\documentclass[epj, final]{svjour}

\usepackage{graphicx}
\usepackage{amssymb}
\usepackage{url}
\begin{document}

\newcommand{\etal}{\textit{et~al.}\
}   
\newcommand{\eg}{\textit{e.g.}\
}   
\newcommand{\ie}{\textit{i.e.}\
}   

\title{Delocalization due to correlations in two-dimensional disordered
systems} 

\author{Gabriel Vasseur \and Dietmar Weinmann\thanks{\emph{e-mail:} Dietmar.Weinmann@ipcms.u-strasbg.fr}}

\institute{Institut de Physique et Chimie des Mat\'eriaux de Strasbourg,\thanks{UMR
7504 (CNRS-ULP)} 23 rue du Loess, BP 43, 67034 Strasbourg Cedex 2, France}

\date{\today}

\abstract{ We study the spectral statistics of interacting spinless fermions in a
two-dimensional disordered lattice. Within a full quantum treatment for
small few-particle-systems, we compute the low-energy many-body states
numerically. While at weak disorder the interactions reduce spectral
correlations and lead to localization, for the case of strong disorder we find
that a moderate Coulomb interaction has a delocalizing effect. In addition, we
observe a non-universal structure in the level-spacing distribution which we
attribute to a mechanism reinforcing spectral correlations taking place in small
systems at strong disorder.
} 

\maketitle

\section{Introduction}

The ongoing miniaturization of electronic devices and the peculiar physics
associated motivate the investigation of systems with reduced dimensionality.
In such systems, the effect of Coulomb interactions is expected to be strong,
and prominent experimental observations of the last decade are thought to be due to
correlation effects.

One such experimental result concerns the magnitude of the persistent currents
in disordered mesoscopic rings \cite{levy1990,chandrasekhar1991}, which are much
greater than theoretical predictions from  approaches neglecting correlations
\cite{prl1991}. The interaction effect seems to be beyond the perturbative
regime, but a full treatment of the realistic situation is out of reach (for a
review see \cite{eckern1995}). Nevertheless, analytical \cite{giamarchi1995} and
numerical \cite{gambetti2002,benenti1999} calculations in 1D and in 2D have
shown that a repulsive interaction can enhance the persistent currents.

As another important example, a metallic behaviour has been observed
\cite{kravchenko1994} (see \cite{kravchenko2004} for a review) in 2D electron
gases at low electronic density, where the ratio $r_s$ of Coulomb to kinetic
energy is large ($> 10$). This cannot be explained by the standard scaling
theory of localization which, neglecting electronic correlations, predicts an
insulating behaviour in 2D for any finite disorder strength \cite{abrahams1979}.
Since interactions become important when the electron density is low (large
$r_s$), they have been suggested to be responsible for the observed metallic
behaviour \cite{benenti1999}. However, a perturbative introduction of the
interaction leads to a reinforcement of the electron localization
\cite{altshuler1980}. This points towards the necessity of treating the
interactions beyond perturbation theory, and several approaches have been
proposed in this direction.

Renormalization group techniques indicate that the interaction permits a
metal-insulator transition in a weakly disordered two-dimensional electron gas
\cite{castellani1998,si1998}. However, despite some well reproduced properties
(such as the effect of a magnetic field), a description of the transition is not
possible and the metallic phase is based on the assumption of a Fermi liquid.

Alternatively, field theory based on the fact that the metallic phase is not a
Fermi liquid (as argued in \cite{dobrosavljevic1997}) has shown that a perfect
metal can be stable in 2D if the interaction is strong enough
\cite{chakravarty1998}.

Furthermore, numerical calculation of the current-current correlation function
using a quantum Monte Carlo approach has shown that interactions change the
behaviour of the conductivity at low temperature from an insulating to a
metallic one \cite{denteneer1999}, as observed in the experiments. 

All these approaches contain approximations. On the contrary, numerical studies
of model systems allow for exact results, although they are limited to
simplified models and small system sizes. Therefore, such studies are able to
provide a complementary view of the mechanisms implied in the interplay between
disorder and interaction. 

In the present work, we numerically investigate interacting spinless fermions
(spin-polarized electrons) in small two-dimensional lattices with disorder,
performing a direct diagonalization of the Hamiltonian.

The limitation in size makes it difficult to vary directly the electronic
density as in experiment. To change the interaction parameter $r_s$, we vary
instead the interaction strength $U$ while keeping
the system size $L$ and the particle number $N$ constant.

\begin{table}
\caption{\label{tab} Summary of the behaviour of the distribution $P(s)$ for
different regimes of interaction $U$ and disorder in small two-dimensional
systems. PWMS stands for Pinned Wigner Molecule Statistics (see text). The
cross-over between the weak and strong interaction regimes represent the main
result of the paper.
}
\begin{center}
\begin{tabular}{lll}
\hline\noalign{\smallskip}
$U$             &    Weak disorder     &   Strong disorder   \\
\noalign{\smallskip}\hline\noalign{\smallskip}
$0$             &    Wigner-Dyson      &   Poisson           \\
$\infty$        &    PWMS              &   PWMS              \\
cross-over      &    monotonic         &   non-monotonic     \\
\noalign{\smallskip}\hline
\end{tabular}
\end{center}
\end{table}

Within this approach, we explore the interaction effects on the ground state
structure and the probability density $P(s)$ of the normalized level spacing $s
= \Delta / {\left\langle\Delta\right\rangle}$, where $\Delta = E_1 - E_0$ is the
energy spacing between the many-body ground-state and the first excited state.
We denote by $\left\langle...\right\rangle$ the average over the ensemble of
disorder configurations.

In the non-interacting case, $\Delta$ is equal to the one-body level spacing at
the Fermi energy. The statistics of these one-body level spacings $P(s)$ has
been extensively studied and found to be an indicator of the metal-insulator
Anderson transition occurring in 3d as a function of the disorder strength
\cite{altshuler1986,shklovskii1993}. In the diffusive regime, $P(s)$ corresponds to
Wigner-Dyson statistics, while the Anderson insulating regime is characterized
by Poisson statistics (table \ref{tab}).

In the opposite limit, at $U=\infty$, the structure of the ground state is
imposed by the Coulomb repulsion, which leads to a Wigner crystal pinned by the
disorder. We show in section \ref{ps} that the resulting distribution
$P_{U=\infty}(s)$ (which we call ``Pinned-Wigner-Molecule Statistics'', PWMS) is
non-universal for finite size systems. 

For weak disorder, we found the dependence of $P(s)$ on $U$ to be consistent
with previous studies: $P(s)$ crosses over smoothly from Wigner-Dyson statistics
to its infinite interaction limit \cite{benenti2000,berkovits2003,katomeris2003}.

Within an approximative method (Configuration Interaction method, starting
from Hartree-Fock orbitals), the strong disorder case has been studied
by Benenti \etal for larger systems \cite{benenti2000a}.
They obtain an interaction-induced transition of $P(s)$ from Poisson
to Wigner-Dyson. Our findings of an increase of spectral correlations by
moderate interaction show that this behaviour persists when electronic
correlations are taken fully into account. 

The non-trivial behaviour of the spectral statistics for strong disorder is
discussed in section \ref{sec_ps}, after the presentation of the model we study
(section \ref{sec_mod}). In section \ref{sec_imipr}, we present results for
the inverse participation ratio of the ground state in the many-body on-site
basis, and we show that the increase of spectral correlations is related to a
delocalizing effect. Our conclusions are discussed in section \ref{sec_ccl},
and a small size effect modifying the shape of $P(s)$ is presented in the
Appendix.

\section{\label{sec_mod}Disorder and Coulomb Interaction}

We consider $N$ spinless fermions on a two-dimensional $L$ by $L$ lattice (in
the following, we concentrate on $N=4$ and $L=6$). We note $M=L^2$ the number of
sites.

The Hamiltonian of the system is $H = H_A + H_U$, where $H_A$ is the standard
Anderson Hamiltonian
\begin{equation}
H_A =  -t \sum_{<i,j>}^{} ( c_i^+ c_j + c_j^+ c_i )
    + \sum_{i=1}^{M} v_i c_i^+ c_i,
\end{equation}
with $c_i^+$ ($c_i$) creating (destroying) an electron on the site $i$. The
first term of $H_A$ allows for hopping between nearest neighbours $<i,j>$ on the
lattice. We take $t=1$, representing then the energy scale. Periodic boundary
conditions are used, leading to a toroidal topology. The second term of $H_A$
models a disorder potential. The $v_i$ are independent random variables
uniformly distributed in $[-W/2;W/2]$, and $W$ is the disorder strength.

The interaction term is chosen to be of the Coulomb form:
\begin{equation}
 H_U
  = \displaystyle{\frac{U}{2} \sum_{ \stackrel{\scriptstyle i,j=1}{\scriptstyle i \neq j} }^{M}
     \frac{c_j^+ c_i^+ c_i \, c_j}{d_{ij}} },
\end{equation}
where $d_{ij}$ is the smallest distance on the torus between the sites $i$ and
$j$ and $U$ is the interaction strength.

This model allows us to study qualitatively the effect of interactions in
disordered systems. For the exact diagonalization of the Hamiltonian $H$, we
have used a routine developed by Simon and Wu \cite{TRLan} based on the Lanczos
algorithm \cite{Lanczos1985}.

\section{\label{sec_ps}Interaction induced many-body level-repulsion}

In this section, we study in detail the probability distribution $P(s)$ of the first
excitation energy whose behavior in different regimes is sketched in table
\ref{tab}.

In the absence of interactions, $P_{U=0}(s)$ corresponds to the Wigner-Dyson
distribution $P_{WD}(s)= \frac{\pi}{2} s\exp(-\frac{\pi}{4}s^2)$ for a diffusive
(metallic) system. For a strongly disordered system (Anderson insulator), its
limiting behavior for $L\to\infty$ is the Poisson distribution
$P_P(s)=\exp(-s)$\footnote{In finite systems, $P_{U=0}(s)$ is distinct from
Poisson even for $W\to\infty$. For the system size we consider ($M=36$),
however, the difference is rather small.}. 

In the strong interaction limit ($U\to\infty$), the electrostatic energy
dominates and the energetically lowest many-body states correspond to periodic
distributions of the electrons (Wigner crystal) which are pinned even by a small
amount of disorder. All Wigner crystals have the same electronic structure, thus
the same Coulomb energy, but differ in their location on the lattice and
therefore in their disorder energy.

In the limit of low densities, the ground state structure approaches the usual
hexagonal Wigner crystal, the number of possible positions on the lattice is
large, and $P_{U=\infty}(s)$ tends to Poisson. For higher densities we need to
take into account commensurability effects. In the studied case, $N$ and $L$ are
such that the nine energetically lowest $U=\infty$ many-body states are
square-shaped Wigner crystals (which we will refer to as ``Wigner molecules'')
differing only by their location on the lattice. These nine Wigner molecules
have different disorder energy, and the first excitation energy is
\begin{equation} \Delta = E_1 - E_0 = \sum_{k=1}^{N} v_{i_k}  - \sum_{k=1}^{N}
v_{j_k}, \end{equation} where the $j_k$ describe the sites occupied by a
particle in the energetically lowest Wigner Molecule (the ground state) and the
$i_k$ the occupied sites in the second Wigner molecule (the first excited
state). Since we only have nine configurations, $P_{U=\infty}(s)$ is not exactly
Poisson. Instead, it is a distribution which is intermediate between
semi-Gauss\footnote{At half filling there are only two different Wigner
crystal configurations. Their energy difference is given by the difference between two
sums of independent random numbers. According to the central limit theorem,
its distribution in the limit of $N\to\infty$ is Gaussian. Since only positive
values for $s$ are considered, the resulting distribution $P(s)$ is called
semi-Gauss.} and Poisson, which we call pinned-Wigner-molecule statistics (PWMS).

 \begin{figure}
\resizebox{0.95\columnwidth}{!}{%
\includegraphics{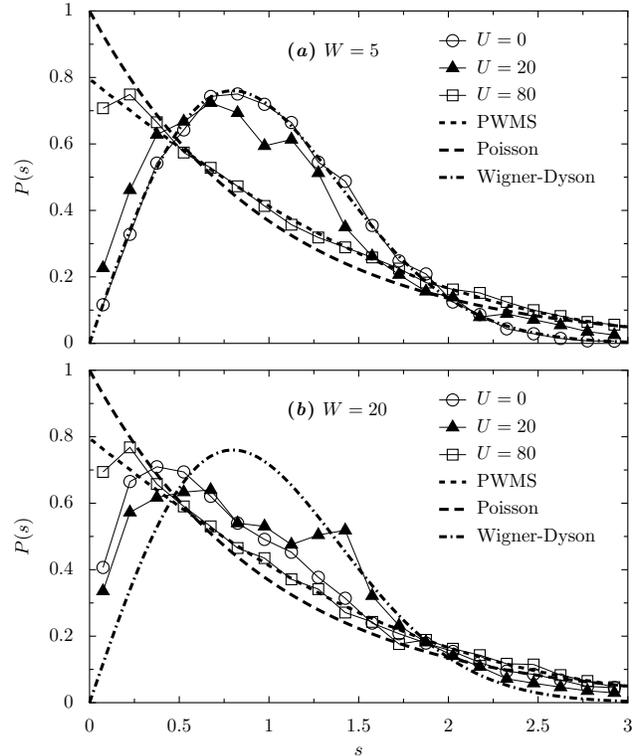}
}
 \caption{Distribution of the lowest many-body excitation for different
 interaction strengths $U$ at $W=5$ (\textit{a}) and $W=20$ (\textit{b}), for 4
 particles on a 6 by 6 lattice. Each curve is obtained from data for 9000
 disorder configurations.} 
 \label{ps} 
 \end{figure}

Figure \ref{ps}a shows $P(s)$ with $W=5$, for three values of $U$. With this disorder
strength, the single-particle localization length is larger than the system
size, so that the motion of non-interacting particles through the sample is
diffusive and $P(s)$ follows Wigner-Dyson statistics at $U=0$.

In this case, we find that the level repulsion decreases with increasing
interaction strength. In this sense, this result is consistent with those of
\cite{benenti2000,berkovits2003,katomeris2003}. The difference is that the
$U=\infty$ limit of $P(s)$ is in fact the PWMS and not Poisson for the
investigated system sizes.

This behaviour is as expected: strong interactions drive the system out of the
diffusive regime by setting up a Wigner Crystal which is pinned by disorder.
Therefore spectral correlations disappear with increasing interaction strength.

Figure \ref{ps}b shows our results for the case of stronger disorder ($W=20$).
This disorder is not strong enough to have a Poissonian $P(s)$ in the
non-interacting case, and hence $P_{U=0}(s)$ is intermediate between Poisson and
Wigner-Dyson. Nevertheless, a single-particle state at the Fermi energy is
localized over about three sites only, meaning that the system is far from the
diffusive regime for $U=0$. 

The main feature is that at this disorder strength, the evolution of $P(s)$ with
the interaction strength is non-monotonic. Obviously, for stronger interactions
the spectral correlations eventually decrease and $P(s)$ approaches its infinite
interaction limit (PWMS). For a moderate $U$, however, the spectral correlations are
increased relative to the non-interacting case (\ie $P(s)$ is more
Wigner-Dyson-like). This is the main result of the present paper. The increase
of spectral correlations could be the precursor of the transition towards
universal correlations found in \cite{benenti2000a}.

The non-monotonic behavior becomes less significant as $W$ increases further. It
is for this reason that we have presented our results for $W=20$. Note that,
with interaction, a peak appears in the $P(s)$ curves. This peak is the
manifestation of a mechanism enhancing spectral correlations in a non-universal
way for small system size. We discuss this mechanism in detail in the
appendix.

\begin{figure}[tbp]
\resizebox{0.95\columnwidth}{!}{%
\includegraphics{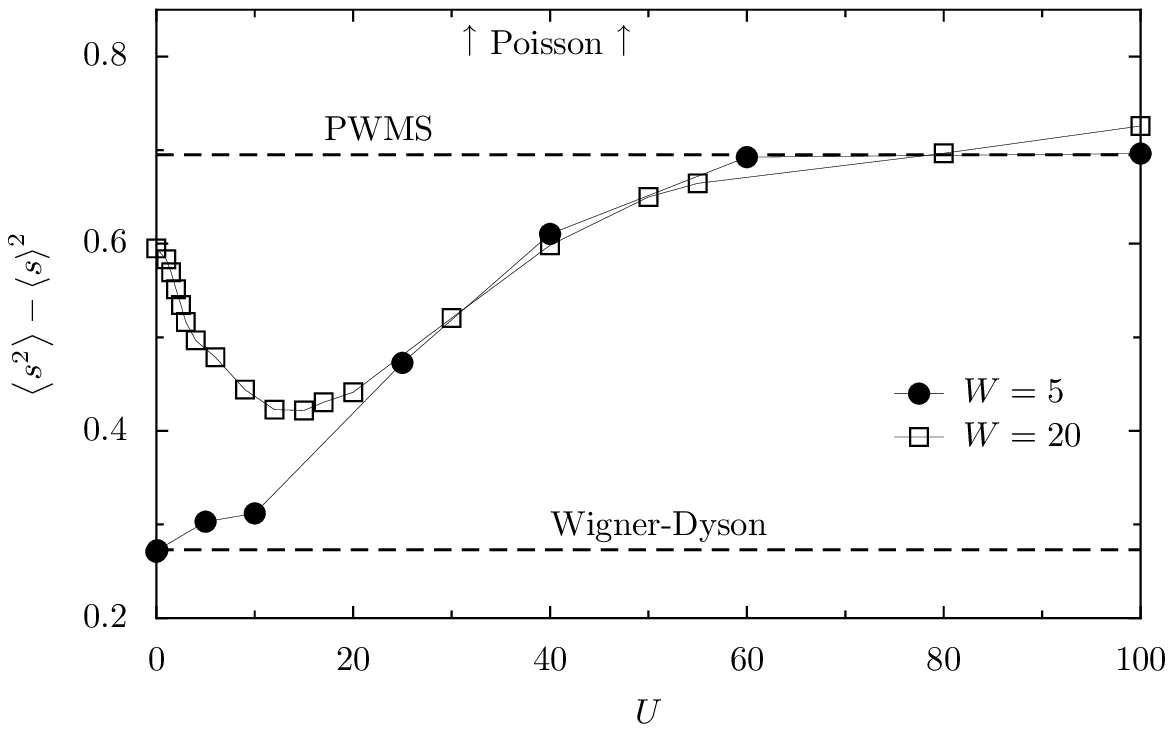}
}
 \caption{The variance of $s$ as a function of $U$ at $W=5$ (filled circles) and $W=20$
 (open squares), for 4 particles on a 6 by 6 lattice. Each point is computed from
 3000 disorder configurations. The statistical error is smaller than the symbol
 size. The $W=20$ infinite $U$ limit PWMS is approached from above at values of
 $U$ which lie outside the scale of the figure. For Poisson, $\left\langle
 s^2\right\rangle - {\left\langle s\right\rangle}^2=1$ (out of the scale of the
 vertical axis).}
 \label{variance} 
 \end{figure}

In order to characterize quantitatively the non-monotonic behaviour exhibited by
$P(s)$, we consider the evolution of the variance of $s$ with $U$. We show in
figure \ref{variance} that this evolution in the cases of
$W=5$ and $W=20$ is very different. At $W=5$ there is a monotonous evolution towards smaller
correlations, while at $W=20$ the variance exhibits a minimum as a function
of $U$ (corresponding to an increase of the spectral correlations).

With a short range interaction, we have found similar results for
$P(s)$ at half filling.
However, away from half filling, configurations not affected by interactions are
connected by hopping matrix elements. Therefore, short range
interactions cannot suppress the mobility of the electrons. As a
consequence, the
variance (or another parameter characterizing the distribution $P(s)$) 
does not reach its PWMS value at strong $U$.

\section{\label{sec_imipr}Delocalization in the many-body on-site basis}

The fact that $P(s)$ approaches the Wigner-Dyson distribution in the presence of
a moderate interaction could be interpreted as a signature of a delocalization
of the electrons. Whereas the link between spectral correlations and
localization is clear for one-particle level statistics ($U=0$), it is less
obvious in the interacting regime. To clarify this interpretation, we have
studied the localization of the ground-state \begin{equation}
\left|g\right\rangle = \sum_{n} {\Psi}_n \left|n\right\rangle \end{equation} in
the many-body on-site basis $\{\left|n\right\rangle\}$, via its inverse
participation ratio \begin{equation} \mathcal{R}^{-1} = \sum_{n}
{\left|{\Psi}_n\right|}^4. \end{equation}

Contrary to $P(s)$, the inverse participation ratio
$\mathcal{R}^{-1}$ depends on the choice of the basis in which it is calculated.
To be allowed to interpret the inverse participation ratio as a measurement of
the electron localization, we have chosen for $\{\left|n\right\rangle\}$ the
Slater determinants
\begin{equation}
c_{i_1}^+ c_{i_2}^+ c_{i_3}^+ c_{i_4}^+ \left|0\right\rangle
\end{equation}
which correspond to the four particles being localized on the lattice sites
$i_1$, $i_2$, $i_3$ and $i_4$ and where $\left|0\right\rangle$ is the empty
lattice state.

If $\left|g\right\rangle$ is given by one of these basis states
$\left|m\right\rangle$, then $\Psi_n = \delta_{n,m}$, and $\mathcal{R}^{-1}=1$.
If, on the other hand, $\left|g\right\rangle$ is a superposition of many
elements of $\{\left|n\right\rangle\}$, then $\mathcal{R}^{-1}$ is very small
compared to unity. Since the basis is built with completely localized
electrons, $\mathcal{R}^{-1}$ can be interpreted as a measurement of the
localization of electrons in the many-body state $\left|g\right\rangle$.

 \begin{figure}[tbp]
\resizebox{0.95\columnwidth}{!}{%
\includegraphics[width=8.5cm]{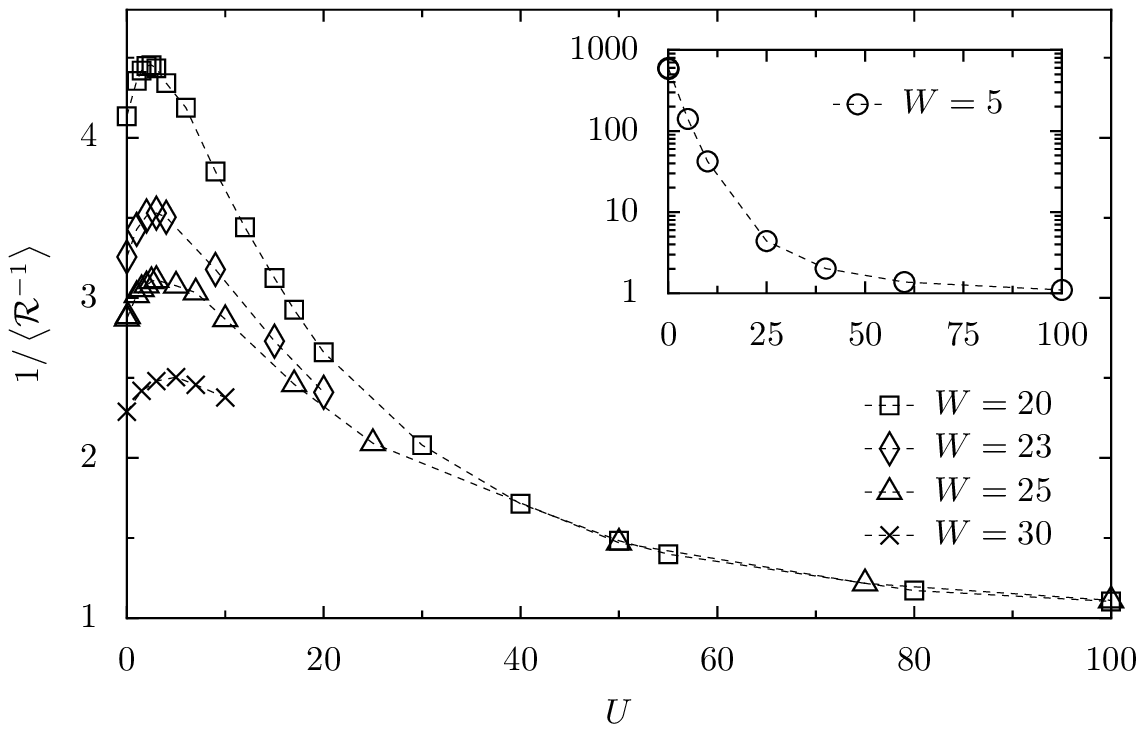}
}
 \caption{Evolution of $1 / \left\langle \mathcal{R}^{-1}\right\rangle$ as a
 function of $U$ for $W=5$ (circles, in the inset), $W=20$ (squares), $W=23$
 (diamonds), $W=25$ (triangles) and $W=30$ (crosses). The statistical error is
 smaller than the symbol size.}
 \label{mipr}
 \end{figure}
Figure \ref{mipr} shows numerical results for the inverse of the average of
$\mathcal{R}^{-1}$. In the case of weak disorder ($W=5$, in the inset), this
quantity is monotonically decreasing with $U$, as expected. The $U=0$ diffusive
situation with delocalized particles is perturbed by moderate interactions which
increase the scattering, reducing both the mobility and
$1/\left\langle\mathcal{R}^{-1}\right\rangle$. In the regime of strong
interactions, the electrons form a Wigner crystal to minimize electrostatic
energy. Since this corresponds to one particular state of the chosen basis,
$\mathcal{R}^{-1}$ decreases to one.

In the regime of strong disorder, the behaviour is very different. At weak
interaction, $ 1 / \left\langle\mathcal{R}^{-1}\right\rangle$ increases with
$U$, which means that interaction has a delocalizing effect. This can be
understood within the following scenario. 

At $U=0$, we have localized one-body wave functions, the disorder being strong
enough to dictate the electronic configuration. For finite $U$, the effect of
the interaction depends on this particular sample-dependent configuration. In
some samples, this electronic structure is close to the one
adapted to interaction (Wigner molecule), therefore the interaction strengthens
the localization. On the other hand, in most samples, the disorder-adapted
electronic configuration is rather different from the Wigner molecule structure.
Therefore, increasing interaction strength induces charge reorganizations (as
proposed in \cite{weinmann2001} in the context of one-dimensional rings) at
particular sample-dependent values of the interaction strength $U_c$. When
$U\simeq U_c$, the competition between interaction and disorder leads to a
ground state which is a superposition of a state adapted to disorder and another
one adapted to interaction. This results in a pronounced delocalization. 

Since $U_c$ is strongly sample-dependent, these delocalizations are smoothed by
the disorder average. Furthermore, we expect that as $W$ increases, the mean
value of $U_c$ also increases and its distribution spreads, consistent with the
behaviour observed in figure \ref{mipr}.
 
\begin{figure}
\resizebox{0.95\columnwidth}{!}{%
\includegraphics[width=8.5cm]{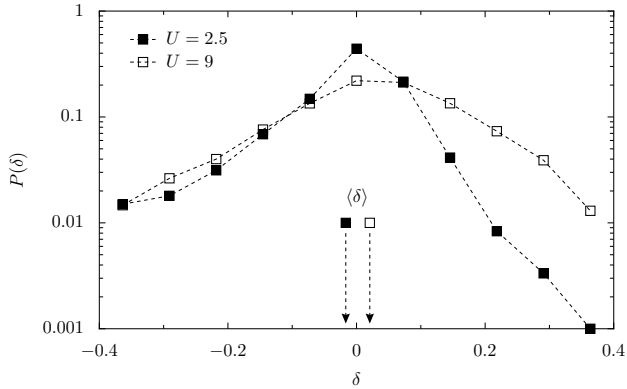}
}
 \caption{Distribution of $\delta$ at
 $W=20$ for $U=2.5$ (filled squares) and $U=9$ (open squares) calculated from 3000
 disorder configurations. The averages are indicated below the curves. The
 statistical error for these averages is smaller than the symbol size.}
 \label{delta}
 \end{figure}

To illustrate these considerations, we can define for each sample the
increase
\begin{equation}
\delta = \mathcal{R}^{-1}(U) - \mathcal{R}^{-1}_0
\end{equation}
of $\mathcal{R}^{-1}$ with respect to the non-interacting value
$\mathcal{R}^{-1}_0$.
Figure \ref{delta} depicts the distribution of this quantity at $W=20$, for two
different values of $U$.
We can see that the most probable value of $\delta$ is close to zero.
Nevertheless, for $U=2.5$ the negative tail of the distribution, corresponding
to strong delocalizations, is more pronounced than the positive one, which
corresponds to strong localizations, resulting in a negative average of
$\delta$. This is not the case when the interaction is strong ($U=9$). We can
conclude that for $W=20$, $U_c$ is more probable to be close to $2.5$ than $9$.

\section{\label{sec_ccl}Conclusion}

We have studied the spectral statistics of interacting spinless
fermions in a two-dimensional disordered system, using exact diagonalization.

We have found that correlations in the statistics of the level spacing between
the many-body ground state and the first excited state are increased by a
moderate interaction when the disorder is strong enough to localize the
one-particle wave functions. The interpretation of this effect as a
delocalization of the electrons has been supported by the study of the inverse
participation ratio for the ground state.

The delocalization effect can be understood as a consequence of a competition
between disorder and interaction for the structure of the ground-state, taking
place at some sample-dependent value of the interaction strength. This
competition results in charge reorganization caused by increasing interaction
from a configuration minimizing the disorder energy to a configuration more
adapted to the interaction.

\begin{acknowledgement}
We acknowledge very useful discussions with and comments
by R. A. Jalabert, P. E. Falloon, G.-L. Ingold, and J.-L. Pichard.
\end{acknowledgement}

\appendix
\section{Appendix: Non-universal level statistics in small systems}

The peak appearing in the $P(s)$ curve of figure \ref{ps}b for $U=20$ is the
manifestation of a mechanism enhancing spectral correlations in a non-universal
way. Since its major ingredient is a competition between disorder and
interaction, this mechanism could be the precursor of what happens in bigger
systems, where other mechanisms could take place at higher order,
resulting in enhanced spectral correlations even in the thermodynamic limit.

The mechanism we want to describe takes place in very small systems when the
mean level spacing $\left\langle\Delta\right\rangle$ is not much smaller than
the hopping $t$. In order to explain this mechanism, we start by considering the
simpler non-interacting situation, before treating the interacting case.

\subsection{Non-interacting case}

Without the hopping $t$, the eigenstates of the system are Slater determinants
of particles localized on single sites, and the distribution of their energies
is Poisson. In the limit of strong disorder, $W\gg t$, the coupling $t$ of
neighbouring sites is typically much smaller than their difference in on-site
energy. Therefore, particles remain localized on single sites, except in samples
for which the energy of the highest occupied site $i$ is almost degenerate with
that of the lowest unoccupied site $j$, and with these two sites being nearest
neighbours.

In those samples, the hopping couples directly two almost degenerate levels,
resulting in a delocalization of one electron over these two sites (as pointed
out in \cite{wobst2003}). 

If the mean spacing $\left\langle\Delta\right\rangle$ is bigger than $t$ (small
size or very big $W$), the two coupled levels can be considered as a two-level
system. Therefore the level spacing is
\begin{equation}
\Delta = \sqrt{{(v_j-v_i)}^2 + 4 t^2}.
\end{equation}
In the case where the sites $i$ and $j$ are not nearest neighbours
$\Delta=v_j-v_i\simeq 0$. As a consequence, the special samples are responsible for the
appearance of a dip and a peak in $P(s)$ at $s=0$ and
$s=2t/\left\langle\Delta\right\rangle$, respectively.

On the other hand, if the mean spacing $\left\langle\Delta\right\rangle$ is
smaller than $t$ (greater size and not too big $W$), typically more than two
levels are coupled and eventually universal random-matrix-theory-like
correlations can arise. Even though at intermediate sizes the spectral
correlations are still greater in the special samples, their weight in the
ensemble rapidly decreases with $L$. In the thermodynamic limit, the anomaly
therefore disappears.

\subsection{Interacting case}

In the absence of hopping ($t=0$), given a quite strong disorder $W$, it is
always possible to find values of $U$ such that the two energetically lowest
many-body states of a given sample are almost degenerate. For example, one can
minimize the interaction energy (with a Wigner crystal) and the other one can be
slightly different, increasing the interaction energy while reducing disorder
energy.

Now, the first two many-body states can in certain samples be connected by only
a single hop of one particle. The probability for such a situation is quite
large since one many-body state is coupled to many others by the hopping of one
of the particles.

In those samples, the introduction of the hopping couples directly two almost
degenerate levels. If the values of $U$ and $W$ are strong enough, the particles
remain localized on individual sites except the one implied in the connection of
the two states. This results in a delocalization of the ground state in the
on-site basis ($\mathcal{R}=2$).

For $\left\langle\Delta\right\rangle>t$ (small size or very big $W$), the two
coupled levels can again be considered as a two-level system, and the
non-universal correlations appear as explained previously.

For $\left\langle\Delta\right\rangle<t$ (greater size and not too big $W$),
typically more levels are coupled and the two-level system approximation breaks
down. As for the non-interacting case, this mechanism disappears in the
thermodynamic limit. However, other mechanisms involving more hoppings can take
place, and the competition between disorder and interaction might still induce
delocalization and eventually stem the spectral correlations seen in
\cite{benenti2000a}.

\bibliography{bibliography,article}
\end{document}